\documentclass[12pt]{iopart}
\pdfoutput=1

\expandafter\let\csname equation*\endcsname\relax
\expandafter\let\csname endequation*\endcsname\relax
\usepackage{amsmath}
\usepackage{amssymb}
\usepackage{hyperref}
\hypersetup{urlcolor=blue, colorlinks=true}  

\usepackage{color}
\usepackage{graphicx}
\usepackage{extarrows}
\usepackage[sort&compress,numbers]{natbib}

\usepackage{bm}
\usepackage{bbm}
\usepackage{subcaption}

\newcommand{\sx}{\sigma^{x}}
\newcommand{\sy}{\sigma^{y}}
\newcommand{\sz}{\sigma^{z}}

\newcommand{\be}{\begin{equation}}
\newcommand{\ee}{\end{equation}}
\newcommand{\bea}{\begin{eqnarray}}
\newcommand{\eea}{\end{eqnarray}}

\newcommand{\ket}[1]{|#1\rangle}
\newcommand{\bra}[1]{\langle#1|}

\newcommand{\one}{1 \hspace{-1.0mm}  {\bf l}}

\newcommand{\Det}{\text{Det}}
\newcommand{\Cov}{\text{Cov}}

\newcommand{\HH}{\mathcal{H}}

\newcommand{\QED}{\hfill\ensuremath{\blacksquare}}

\DeclareRobustCommand\openzero{\leavevmode\hbox{0\kern-.55em0}}

\renewcommand{\Im}{\text{\bf Im}}
\renewcommand{\Re}{\text{\bf Re}}

\begin{document}

\title{On quantumness in multi-parameter quantum estimation}

\author{Angelo Carollo$^{1,2}$, Bernardo Spagnolo$^{1,2,3}$, Alexander A. Dubkov$^{2}$ and Davide Valenti$^{1,4}$}
\address{$^{1}$Dipartimento di Fisica e Chimica "Emilio Segr\`{e}", Group of Interdisciplinary Theoretical Physics, Universit\`{a} di Palermo, Viale delle Scienze, Ed. 18, I-90128 Palermo, Italy}
\address{$^{2}$Radiophysics Department, National Research Lobachevsky State University of Nizhni Novgorod, 23 Gagarin Avenue, Nizhni Novgorod 603950, Russia}
\address{$^{3}$Istituto Nazionale di Fisica Nucleare, Sezione di Catania, Via S. Sofia 64, I-90123 Catania, Italy}
\address{$^{4}$Istituto di Biomedicina ed Immunologia Molecolare (IBIM) "Alberto Monroy", CNR, Via Ugo La Malfa 153, I-90146 Palermo, Italy}

\ead{angelo.carollo@unipa.it}

\begin{abstract}
{In this article we derive a measure of quantumness in quantum multi-parameter estimation problems. We can show that the ratio between the mean 
Uhlmann Curvature and the Fisher Information provides a figure of merit which estimates the amount of incompatibility arising from the quantum nature 
of the underlying physical system. This ratio accounts for the discrepancy between the attainable precision in the simultaneous estimation of multiple
parameters and the precision predicted by the Cram\'er-Rao bound. As a testbed for this concept, we consider a quantum many-body system in thermal equilibrium, and explore the quantum compatibility of the model across its phase diagram.}
\end{abstract}



\vspace{2pc}

\submitto{\JSTAT}

\maketitle

\section{Introduction}
Estimation theory is the discipline that studies the accuracy by which a given set of physical parameters can be evaluated. When the parameters to be estimated 
belongs to an underlying quantum physical system one falls in the realm of quantum estimation theory, or quantum metrology~\cite{Helstrom1976}. Quantum parameter 
estimation finds applications in a wide variety of fields, from fundamental physics~\cite{Udem2002,Katori2011,Giovannetti2004,Aspachs2010,Ahmadi2014}, to gravitational 
wave interferometry~\cite{Schnabel2010,Aasi2013}, thermometry~\cite{Correa2015,DePasquale2016}, spectroscopy~\cite{Schmitt2017,Boss2017}, 
imaging~\cite{Tsang2016,Nair2016,Lupo2016}, to name a few. Exploiting remarkable features of quantum systems as probes may give an edge over the accuracy of 
classical parameter estimation. Exploring this possibility plays a pivotal role in the current development of quantum technology~\cite{Caves1981,Huelga1997,Giovannetti2006,Paris2009,Giovannetti2011,Toth2014,Szczykulska2016,Pezze2016,Nichols2018,Braun2018}. In multi-parameter 
quantum estimation protocols, several variables are simultaneously evaluated, in a way which may outperform individual estimation strategies with equivalent 
resources~\cite{Humphreys2013,Baumgratz2016}, thereby motivating the use of such protocols in a variety of diverse contexts~\cite{Humphreys2013,Baumgratz2016,Pezze2017,Apellaniz2018}.

The use of peculiar quantum many-body states as probes in quantum metrology can enhance the accuracy in parameter estimation~\cite{Zanardi2008,Braun2018}. Conversely, 
one may think of using quantum metrological tools in the study and characterisation of many-body systems. Noteworthy instances of many-body quantum systems are those 
experiencing quantum phase transitions. Indeed, quantum parameter estimation, with its intimate relation with geometric information, provides a novel and promising approach to 
investigate equilibrium~\cite{Carollo2005,Zhu2006,Hamma2006,Zanardi2006,CamposVenuti2007,CamposVenuti2008,Zanardi2007,Zanardi2007a,Garnerone2009a,Rezakhani2010,Bascone2019} 
and out-of-equilibrium~\cite{Magazzu2015,Magazzu2016,Guarcello2015,Spagnolo2015,Spagnolo2017,Spagnolo2018,Valenti2018,Spagnolo2018a} quantum critical phenomena~\cite{Banchi2014,Marzolino2014,Kolodrubetz2013,Carollo2018,Carollo2018a,Marzolino2017}.

The solution of a parameter estimation problem amounts to find an estimator, {\em i.e} a mapping $\hat{\bm{\lambda}}=\hat{\bm{\lambda}} (x_1,x_2,...)$ from the set 
$\chi$ of measurement outcomes into the space of parameters $\bm\lambda \in\mathcal{M}$.  Optimal unbiased estimators in classical estimation theory are those 
saturating the Cram\'er-Rao (CR) inequality
\begin{align}\label{eq:CCRB} 
\Cov_{\bm\lambda}[\hat{\bm{\lambda}}] \geq J^{c} (\bm \lambda)^{-1} 
\end{align}
which poses a lower bound 
on the mean square error $\Cov_{\lambda} [\hat{\bm{\lambda}}]_{\mu\nu} =
E_{\lambda} [(\hat{\lambda} - \lambda)_\mu(\hat{\lambda}-\lambda)_\nu]$ in terms of the Fisher information (FI) \be J^{c}_{\mu\nu}(\bm\lambda) = \int_\chi 
d\hat{\bm{\lambda}}(x)\, p(\hat{\bm{\lambda}}|\lambda) \partial_\mu \log p(\hat{\bm{\lambda}}|\bm\lambda) \partial_\nu \log p(\hat{\bm{\lambda}}|\bm\lambda)\:. \ee
The expression~(\ref{eq:CCRB}) should be understood as a matrix inequality. In general, one writes
\[
\tr(W\Cov_{\bm\lambda}[\hat{\bm{\lambda}}] )\ge\tr(W J^{c}(\bm\lambda)^{-1}),
\]
where $W$ is a given positive definite cost matrix, which allows the uncertainty cost of different parameters to be weighed.

In the classical estimation problem, both in the single parameter case, and in the multi-parameter one, the CR bound can be attained in the asymptotic limit of an infinite 
number of experiment repetitions using the maximum likelihood estimator \cite{Kay1993}. However, an interesting difference between single and multi-parameter 
metrology arises due to correlations between the variables. Indeed, it may well happen that the off-diagonal elements of the Fisher information matrix are
non-vanishing. Hence, there are statistical correlations between the parameter estimators. In a protocol in which all variables but $\bm\lambda_{\mu}$ are precisely 
known, the single-parameter CR bound implies that the best attainable accuracy in estimating $\bm\lambda_{\mu}$ is given by $\text{Var}(\hat{\bm\lambda}) \geq 1/J^{c}_{\mu\mu}$. 
However, in a scenario in which all parameters are simultaneously estimated, one finds that the ultimate precision is lower bounded by $\text{Var}(\hat{\bm\lambda}) \geq 
(J^{c}(\bm\lambda)^{-1})_{\mu\mu}$. A straightforward calculation shows that, for positive-definite matrices, $(J^{c}(\bm\lambda)^{-1})_{\mu\mu} \geq 1/J^{c}(\bm\lambda)_{\mu\mu}$, 
where the inequality is saturated only for vanishing off-diagonal elements. In the limit of a large number of experiment repetitions the CR bound is attainable. This means 
that the equivalence between the simultaneous and the individual protocols in the asymptotic limit holds only if the Fisher information is a diagonal matrix, i.e. if the estimators 
are not correlated~\cite{Cox1987}.

Obviously, any given real positive definite matrix can be transformed via an orthogonal rotation into a diagonal matrix. This clearly implies that there is always a combination 
of the parameters for which the Fisher information matrix is diagonal. However, this choice should be contrasted with the physical opportunity of performing such a rotation, as the 
choice of the parameters we are interested in may arise as a result of physical considerations and in this sense determine a preference in a specific basis.

The underlying quantities used in the derivation of classical Fisher information are parameter-dependent probability-distributions of the data, whereas the objects involved in 
the quantum estimation problems are density operators $\rho(\bm\lambda)$  labelled by ${\bm\lambda\cal \in M}$. Hence, a further difficulty of quantum estimation protocols 
is devising the optimal measurement strategy which gathers from the density matrix the greatest amount of information on the labelling parameters. For single parameter estimation, 
the solution is quite straightforward. If one maximises the classical Fisher information over all possible quantum measurements, the result is the so-called quantum Fisher information 
(QFI). The key object involved in the calculation of the QFI is the so-called \emph{symmetric logarithmic derivative} (SLD), $L$, a Hermitian operator which is implicitly defined as 
the solution of the equation 
\be\label{eq:SLD} \frac{d \rho}{d\lambda} = \frac12 \left ( \rho  L
+ L \rho  \right )\:. 
\ee 
The QFI can be calculated using the formula
\begin{align}
 J =\Tr (\rho L^{2}), 
\end{align} 
It turns out that choosing the projective measurement in the eigenbasis of the SLD defines the optimal 
strategy which yields  the FI equal to the QFI. Thus, the QFI defines the ultimate attainable accuracy in the parameter estimation of density matrices $\rho(\bm\lambda)$, in the 
limit of an infinite number of experimental outcomes. The straightforward generalization of the above arguments to the multi-parameter case leads to the so-called multiparameter 
quantum CR bound~\cite{Helstrom1976,Holevo2011,Paris2009}, that reads
\begin{equation}\label{eq:CRB}
\tr(W \Cov(\hat{\bm{\lambda}}))\ge\tr( W J^{-1}),
\end{equation}
where
\begin{align}\label{FI}
J_{\mu\nu}=\frac{1}{2}\Tr\rho\{L_{\mu},L_{\nu}\},
\end{align}
is the quantum Fisher information matrix (QFIM), $W$ is the cost matrix, and $L_{\mu}$ is the SLD implicitly defined by~(\ref{eq:SLD}), with $\rho$ derived with respect to the 
parameter $\lambda_{\mu}$.\\ 
\indent For single parameter estimation, the quantum Cram\'er-Rao bound~(\ref{eq:CRB}) can always be saturated by a suitable optimal positive-operator valued measure 
(POVM). However, in a multi-parameter scenario this is not always the case and the above inequality cannot always be attained. This is due to the non-commutativity of 
measurements associated to independent parameters. It turns out that, within a relatively general setting, known as 
\emph{quantum local asymptotic normality}~\cite{Hayashi2008,Kahn2009,Gill2013,Yamagata2013}, the multi-parameter quantum Cram\'er-Rao bound~(\ref{eq:CRB}) is
attainable iff~\cite{Ragy2016}
\begin{align}\label{IncCond}
\mathcal{U}_{\mu\nu}= 0,
\qquad\forall \mu,\nu,
\end{align}
where
\begin{align}\label{MUC}
\mathcal{U}_{\mu\nu}:=-\frac{i}{4}\Tr\rho [L_{\mu},L_{\nu} ].
\end{align}
The Eq.~\eqref{IncCond} is known as \emph{compatibility condition}~\cite{Ragy2016}. In the context of quantum information geometry, and quantum holonomies of mixed states, 
$\mathcal{U}_{\mu\nu}$ is known as mean Uhlmann curvature (MUC)~\cite{Carollo2018,Carollo2018a,Leonforte2019,Bascone2019,Leonforte2019a,Bascone2019a}.\\
\indent From a metrological point of view, $\mathcal{U}_{\mu\nu}$ marks the \emph{incompatibility} between $\lambda_{\mu}$ and $\lambda_{\nu}$, where such an incompatibility arises from the inherent quantum nature of the underlying physical system.\\
\indent In this work, we show that for an \emph{N-parameter estimation model}, the deviation of the \emph{attainable} multi-parameter 
bound from the Cram\'er-Rao bound can be estimated by the quantity 
\begin{align}
R:=|| 2i\, J^{-1}\mathcal{U}||_\infty
\end{align}
where $||.||_\infty$ is the largest eigenvalue of a matrix, and we find
\begin{align}\label{UBound}
0\le R\le 1.
\end{align}
Indeed, $R$ provides a figure of merit which measures the \emph{amount of incompatibility} within a parameter estimation model. The lower limit condition, $R=0$, is equivalent 
to the compatibility condition, Eq.~\eqref{MUC}. On the other hand, when the upper bound of Eq.~\eqref{UBound} is saturated, i.e. $R=1$, it maximizes the discrepancy between 
the CR bound, that could be attained in an analogous classical multi-parameter estimation problem, and the actual multi-parameter quantum CR bound. In this sense, this bound 
marks the \emph{condition of maximal incompatibility}. When this condition is met, the indeterminacy arising from the quantum nature of the estimation problem reaches the order 
of $||J^{-1}||_\infty$, i.e. the same order of magnitude of the Cram\'er-Rao bound~(\ref{eq:CRB}). In other words, this implies that the indeterminacy due to the quantum incompatibility 
arises at an order of magnitude which cannot be neglected.

This is particularly relevant, considering that the scope of optimal schemes is minimising the parameter estimation error. This can only be done by designing strategies which strive 
for the highest possible rate of growth of $J(n)$ with the number $n$ of available resources. When the condition~$R=1$ of maximal incompatibility holds, it implies that the quantum 
indeterminacy in the parameter estimation problem remains relevant even in the asymptotic limit $n\to\infty$.

\section{Multi-parameter Incompatibility: a Measure of Quantumness}\label{sec:quant} Unlike the single parameter case, in the multi-parameter scenario the QFI CR bound cannot 
always be saturated. Intuitively, this is due to the incompatibility of the optimal measurements for different parameters. A sufficient condition for the saturation is indeed $[L_{\mu},L_{\nu}]=0$, 
which is however not a necessary condition. Within the comprehensive framework of quantum local asymptotic normality (QLAN)~\cite{Hayashi2008,Kahn2009,Gill2013,Yamagata2013}, 
a necessary and sufficient condition for the saturation of the multi-parameter CRB is given by $\mathcal{U}_{\mu\nu}=0$ for all $\mu$ and $\nu$~\cite{Ragy2016}. \\
\indent Here, we show explicitly  that the ratio between $\mathcal{U}_{\mu\nu}$ and $J_{\mu\nu}$ provides a figure of merit for the discrepancy between an attainable multi-parameter 
bound and the single parameter CRB quantified by $J^{-1}$. We will confine ourself to the broad framework of QLAN, in which the \emph{attainable} multi-parameter bound is given by 
the so called Holevo Cram\'er-Rao bound (HCRB)~\cite{Helstrom1976,Holevo2011,Paris2009}. For a $N$-parameter model, the HCRB can be expressed as~\cite{Hayashi2008}
\begin{equation}
\tr(W \Cov(\hat{\bm{\lambda}}))\ge C_{H}(W),
\end{equation}
where
\begin{equation}\label{CH}
C_{H}(W):=\min_{\{X_{\mu}\}}\{\tr (W \Re Z) +||\sqrt{W} \Im Z\sqrt{W}||_{1}\}.
\end{equation}
The $N\times N$ Hermitian matrix $Z$ is defined as 
\begin{align}\label{defZ} Z_{\mu\nu}:=\Tr (\rho X_{\mu}X_{\nu})
\end{align}
where $\{X_{\mu}\}$ is an array of $N$ Hermitian operators on $\HH$ satisfying the unbiasedness conditions $\Tr(\rho X_{\mu})=0$, $\forall \mu$ and 
$\Tr (X_{\mu} \partial_{\nu}\rho)=\frac{1}{2}\Tr \rho \{X_{\mu}, L_{\nu}\}=\delta_{\mu\nu}$, $\forall \mu,\nu$, and $||B||_{1}$ denotes the sum of all singular 
values of $B$. 
If one chooses for $\{X_{\mu}\}$ the array of operators $\tilde{X}_{\mu}:=\sum_{\nu} [J^{-1}]_{\mu\nu} L_{\nu}$, it yields
\begin{equation}\label{Z}
Z\to\tilde{Z}:=J^{-1} I  J^{-1} =  J^{-1} - i 2 J^{-1} \mathcal{U} J^{-1},
\end{equation}
where $I$ is the matrix of elements $I_{\mu\nu}:=\Tr{(\rho L_{\mu} L_{\nu})}$, and $\mathcal{U}$ is the skew-symmetric real matrix 
$\mathcal{U}_{\mu\nu}=\frac{i}{4}\Tr(\rho[L_{\mu},L_{\nu}])$. If one indicates by $\mathcal{D}(W):=C_{H}(W) - \tr{W J^{-1}}$  the discrepancy between the 
attainable multi-parameter HCRB and the CRB, one can write the following bounds
\begin{equation}\label{R}
0 \le \mathcal{D}(W)\le \tr{(W J^{-1})} R,
\end{equation}
where 
\begin{align}\label{def:R}
R:=||2i \mathcal{U}J^{-1}||_\infty , 
\end{align}
and the first inequality is saturated iff $\mathcal{U}=0$~\cite{Ragy2016}.

One can show that
\begin{align}\label{R1}
0 \le R \le 1.
\end{align}
When the upper bound~\eqref{R1} is saturated, i.e. the condition $R=1$ is met, it implies that
\begin{equation}\label{AbsIneq}
\mathcal{D}(W) \simeq \tr(W J^{-1}),
\end{equation}
which means that the discrepancy $\mathcal{D}(W)$ reaches the same order of magnitude of the CR bound itself. This limit marks the \emph{condition of 
maximal incompatibility} for the two-parameter estimation problem, arising from the quantum nature of the underlying system. In the opposite limit $R=0$, 
the parameter model is \emph{compatible}, and the discrepancy between the quantum CR bound and its classical counterpart vanishes. Therefore, $R$ 
provides a figure of merit which quantifies the quantum contribution to the indeterminacy of multi-parameter estimations.\\

\emph{Proof of Eq.~\eqref{R}.}
The Eq.~\eqref{R} is justified by the following chain of inequalities
\begin{align}\label{R3}
\mathcal{D}(W)\leq 2||\sqrt{W}\, J^{-1} \mathcal{U} J^{-1}\sqrt{W}||_{1}\le 2||J^{-1/2} W J^{-1}\,  \mathcal{U} J^{-1/2}||_{1}\le\\ \le 2\,||J^{-1/2} W J^{-1/2}||_{1}\,\,\, ||J^{-1/2} \mathcal{U}J^{-1/2}||_{\infty}=\tr{(WJ^{-1})}\,111R.\nonumber
\end{align}
If one takes $A=\sqrt{W}\, J^{-1} \mathcal{U} J^{-1/2}$ and $B=J^{-1/2}\sqrt{W}$, the second inequality follows from Proposision IX.I.1 of Ref.~\cite{Bhatia1997}, which states that 
for any matrix $A$ and $B$ with $AB$ normal, 
\begin{align}
||AB||_{1}\le ||BA||_{1},
\end{align}
and indeed, $AB=\sqrt{W}\, J^{-1} \mathcal{U} J^{-1}\sqrt{W}$ is skew-symmetric. The third inequality of Eq.~\eqref{R} is an application of H\"older's inequality for Shatten-p norms. The last equality of Eq.~\eqref{R} follows from the positive semi-definiteness of $J^{-1/2} W J^{-1/2}=(W^{-1/2} J^{-1/2})^\dag (W^{-1/2} J^{-1/2})$ and the cyclic property of the trace, from which $||J^{-1/2} W J^{-1/2}||_{1}=\tr(J^{-1/2} W J^{-1/2})=\tr(W J^{-1})$.  
Finally, $J^{-1} \mathcal{U}$ is a diagonalisable matrix with the same 
eigenvalues of $J^{-1/2} \mathcal{U} J^{-1/2}$. Indeed, if $J^{-1/2} \mathcal{U} J^{-1/2}=U^\dag D U$, with $D$ diagonal, then $J^{-1} \mathcal{U}=S^{-1}DS$, 
where $S=U J^{1/2}$. Hence, $R:=||i 2 J^{-1}\mathcal{U}||_\infty=2|| J^{-1/2} \mathcal{U} J^{-1/2}||_\infty$. \QED\\

\emph{Proof of Eq.~\eqref{R1}.}
The lower bound comes straightforwardly from Eq.~\eqref{R}. For the upper bound, notice that $Z=\{Z_{\mu\nu}\}$ in Eq.~\eqref{defZ} is a positive semi-definite 
matrix, since $\forall \vec{a}=\{a_\mu\}_{\mu=1}^N\in\mathbb{C}^N$, $\vec{a}^\dag\cdot Z\cdot \vec{a} = \Tr(\rho A^\dag A)\ge 0$, with $A:=\sum_\mu a_\mu X_\mu$. 
Then, from Eq.~\eqref{Z}
\begin{equation}\label{ZR}
J^{1/2}\tilde{Z}J^{1/2}:= \one - i 2 J^{-1/2} \mathcal{U} J^{-1/2}\ge0.
\end{equation}
Since $i 2 J^{-1/2} \mathcal{U} J^{-1/2}$ is a skew-symmetric Hermitian matrix, its eigenvalues are either zero or real numbers that occur in $\pm$ pairs. Then, 
from Eq.~\eqref{ZR} we deduce that these eigenvalues lie within the interval $\{-1,1\}$. Moreover, from the proof of Eq.~\eqref{R} above, $R:=||i 2 J^{-1}\mathcal{U}||_\infty=||i 2 J^{-1/2} \mathcal{U} J^{-1/2}||_\infty\le 1$. \QED
\\\\
For the special case of a two-parameter model, in the eigenbasis of $J$, with eigenvalues $j_{1}$ and $j_{2}$, it holds
\begin{equation}
2iJ^{-1}\mathcal{U}=\left(\begin{array}{cc}j_{1}^{-1}&0 \\0 &
j_{2}^{-1}\end{array}\right)\left(\begin{array}{cc} 0
&\mathcal{U}_{12} \\-\mathcal{U}_{1 2} & 0\end{array}\right)
 = \left(\begin{array}{cc} 0
&2i\frac{\mathcal{U}_{1 2}}{ j_1} \\-2i\frac{\mathcal{U}_{1 2}}{j_2} & 0\end{array}\right).
\end{equation}
It follows that
\begin{equation}\label{AbsIneq}
R=||2i J^{-1} \mathcal{U}||_{\infty}= \sqrt{\frac{\Det \,2 \mathcal{U}}{\Det{J}}}.
\end{equation}
Hence, $\sqrt{\Det \,2 \mathcal{U}/{\Det{J}}}$ provides a figure of merit which measures the \emph{amount of incompatibility} between two independent parameters 
in a quantum \emph{two-parameter} model. \\
\indent For self-adjoint operators $B_{1},\dots,B_{N}$, the Schrodinger-Robertson's uncertainty principle is the inequality~\cite{Robertson1929}
\begin{equation}
\Det\left[\frac{1}{2} \Tr \rho \{B_{\mu},B_{\nu}\}\right]_{\mu,\nu=1}^{N}\ge\Det\left[-\frac{i}{2} \Tr\rho[B_{\mu},B_{\nu}]\right]_{\mu,\nu=1}^{N},
\end{equation}
which, applied to the SLD $L_{\mu}$'s, yields
\begin{equation}\label{DetIneq}
\Det J \ge \Det 2\, \mathcal{U}.
\end{equation}
For $N=2$, when the inequality~(\ref{DetIneq}) \label{AbsIneq} is equivalent to the upper-bound of Eq.~\eqref{R1}, and if saturated, it implies the 
\emph{condition of maximal incompatibility} for the two-parameter estimation problem.\\
\section{Quantum Fisher Information and Incompatibility in Thermal States}
As an application to the above consideration let's consider the rather general context of a thermal state as quantum probe. We will consider the typical scenario in which the parameters of a Hamiltonian of quantum system may vary, and the (possibily unique) thermal equilibrium state changes accordingly. Noteworthy instances are many-body systems in which external control parameters are manipulated and the state of the system changes and a thermal or quantum phase transition may occur. If we consider a generic canonical equilibrium state, $\rho=e^{-i\beta H}/Z$, where $H$ is the parameter dependent Hamiltonian, $E_i$ and $\ket{i}$  are the corresponding eigenvalues and eigenstates, $\beta=1/k_B T$ is the inverse temperature, $p_i=e^{-i\beta E_i}/Z$ and $Z=\sum_i e^{-i\beta E_i}$ is the partition function. Then the QFIM can be expressed as $J=J^c+J^q$ with
\begin{align}
J^c_{\mu \nu}&=\sum_i \frac{\partial_\mu p_i \partial_\nu p_i}{p_i},\label{JcP}\\
J^q_{\mu \nu}&= 2\sum_{ij}^{ E_i\neq E_j} \frac{(p_i - p_j)^2}{(p_i + p_j)} \frac{(\partial_\mu H)_{ij}(\partial_\nu H)_{ji}}{(E_j - E_i)^2},\label{JqP}
\end{align}
where $J^c$ and $J^q$ are the so called "classical" and "quantum" parts of the QFIM, respectively, $(\partial_\nu H)_{ij}:=\bra{i}\partial_\nu H\ket{j}$. The classical part, $J^c$, accounts for the changes in the eigenvalues of the density matrix $\rho$ due to the changes in the Hamiltonian parameters, $\lambda_mu$'s, and it can be related to the isothermal static susceptibility of classical statistical physics~\cite{Prokopenko2011}. The quantum contribution, $J^q$, on the other hand, is concerned with the changes in the eigenstates of the $H$. The incompatibility term, i.e. the mean Uhlmann curvature, reads as follows  
\begin{align}\label{UP}
\mathcal{U}_{\mu \nu}&=i \sum_{ij}^{ E_i\neq E_j} \frac{(p_i - p_j)^3}{(p_i + p_j)^2} \frac{(\partial_\mu H)_{ij}(\partial_\nu H)_{ji}}{(E_j - E_i)^2}.
\end{align}    
Both $J^q$ and $\mathcal{U}$ can be cast in a useful form~\cite{Hauke2015,CamposVenuti2007,Leonforte2019}
\begin{align}
J^q&=- \frac{4}{ \pi} \int_{0}^{+\infty} \frac{d\omega}{\omega^2} \tanh \left( \frac{\omega \beta}{2} \right) \chi^+(\omega),\\
\mathcal{U}&= -\frac{2i}{ \pi} \int_{0}^{+\infty} \frac{d\omega}{\omega^2} \tanh^2 \left( \frac{\omega \beta}{2} \right) \chi^-(\omega),
\end{align}
where $\chi^\pm_{\mu\nu}(\omega):=\frac{\chi_{\mu\nu}''(\omega)\pm\chi_{\nu\mu}''(\omega)}{2}$ are symmetric and skew-symmetric parts of the dissipative dynamical susceptibility tensor
$\chi_{\mu \nu}''(\omega):=-i\pi\sum_{i j} (\partial_{\mu}H)_{i j} (\partial_{\nu}H)_{j i} (p_i - p_j) \delta (\omega - \omega_{ji})$, and $\omega_{ij}:=E_i-E_j$. Making use of the fluctuation-dissipation theorem $
\chi_{\mu \nu}''(\omega)=-\frac{1}{2}[1-e^{-\omega \beta}] S_{\mu \nu}(\omega)$ yields
\begin{align}
J^q&=\frac{2}{ \pi} \int_{-\infty}^{+\infty} \frac{d\omega}{\omega^2} \tanh^2 \left( \frac{\omega \beta}{2} \right) S^+(\omega)\label{JS}\\
\mathcal{U}&=\frac{i}{ \pi}  \int_{-\infty}^{+\infty} \frac{d\omega}{\omega^2} \tanh^2 \left( \frac{\omega \beta}{2} \right) S^-(\omega)\label{US}
\end{align}
where $S^\pm_{\mu\nu}(\omega):=\frac{S_{\mu\nu}(\omega)\pm S_{\nu\mu}(\omega)}{2}$ are the symmetric and skew-symmetric parts of the dynamical structure factor $S_{\mu\nu}(\omega):= \int_{-\infty}^{\infty} dt e^{i\omega t} \langle\partial_\mu H(t) \partial_\nu H(0)\rangle =\pi \sum_{ij} p_i(\partial_{\mu}H)_{i j} (\partial_{\nu}H)_{j i} \delta(\omega - \omega_{ji})$, and $\partial_{\mu}H(t):=e^{iHt} \partial_{\mu}H e^{-iHt}$.
The above expressions allow one to relate both the Fisher information matrix and the MUC to experimentally accessible quantities such as the dynamical response of the system to perturbations $\partial_\mu H$. These expressions provide also the means to study the quantum critical scaling of the expression $R$ above.

\subsection{Quantum Critical Scaling}\label{sec:Critical}
In this subsection we will assume that the system under scrutiny is a many-body system, on a $d$-dimensional lattice of length $\ell$ and spacing $a$, and that the operators $\partial_\mu H$ are local ones, i.e. $\partial_\mu H=\sum_r O_\mu (r)$. It is convenient to deal with intensive quantities, so we will scale each quantity by the size of the system, i.e. $J\to J/\ell^d$, $\mathcal{U}\to \mathcal{U}/\ell^d$. 

The scaling behaviour of $J^q$ and $\mathcal{U}$ close to a critical point $ h= h^c$ follows from standard scaling hypothesis~\cite{Mussardo2010}. Consider a scaling transformation $a\to \alpha a$, with $\alpha>0$. Close to the critical point, all local operators can be expanded in the basis of the local scaling operators. Assume that $O_\mu$ are such operators, one then has $O_\mu\to \alpha^{-\Delta_\mu}O_\mu$ under rescaling, where $\Delta_\mu$ is the scaling dimension of $O_\mu$. Temperature and frequencies rescale in the same way, i.e. $T\to \alpha^z T$ and $\omega\to\omega^z$, where $z$ is the dynamical critical exponent. By looking at the Eqs.~\eqref{JS} and~\eqref{US}, one realises that $J^q/\ell^d$ and $\mathcal{U}/\ell^d$ scale as $\frac{1}{ \ell^d} \int_{-\infty}^{+\infty} \frac{d\omega}{\omega^2} S^\pm(\omega)$, due to the fact that the remaining terms in the integrals in Eqs..~\eqref{JS} and~\eqref{US} are scale invariant. Since the symmetric  structure factors behaves as $\int_{-\infty}^{\infty} d\omega S_{\mu\nu}^+ \sim \langle \{ O_\mu,O_\nu\} \rangle$, then 
\begin{align}
J^q_{\mu\nu}\ell^d \to J^q_{\mu\nu}/\ell^d \alpha^{-\Delta^J_{\mu\nu}}\qquad\textrm{with }\quad \Delta^J_{\mu\nu}= \Delta_\mu+\Delta_\nu -d -2z.
\end{align}
Similarly, the skew-symmetric  structure factors behave as $\int_{-\infty}^{\infty} d\omega S_{\mu\nu}^- \sim \langle [O_\mu,O_\nu] \rangle$, where the commutator is expected to scale with a scaling dimension $\Delta^-_{\mu\nu}\lesssim\Delta_\mu+\Delta_\nu$.
Thus, one finds
\begin{align}
\mathcal{U}_{\mu\nu}/\ell^d\to \mathcal{U}_{\mu\nu}/\ell^d \alpha^{-\Delta^U_{\mu\nu}}\qquad\textrm{with }\quad \Delta^U_{\mu\nu}= \Delta^-_{\mu\nu} -d -2z.
\end{align}
In the limit of $T$ sufficiently small, where one can neglect the classical part of the QFIM, and in the hypothesis in which both $\Delta^J>0$ and $\Delta^U>0$, so that scaling is dominated by their universal behaviours, one then expects, according to definition~\eqref{def:R}, that
\begin{align}
R\to R \alpha ^{-\Delta_R} \qquad\textrm{with }\quad \Delta_R = \Delta^-_{\mu\nu} -\Delta_\mu-\Delta_\nu\leq 0.
\end{align}
In addition to this, in certain circumstances, such a scaling ansatz is modified by logarithmic multiplicative corrections, such as
\begin{align}
R\to R \,\,(\log{\alpha})^{\tilde{\Delta}_R}\,\,\alpha ^{-\Delta_R}.
\end{align}
For small enough temperature, and large system sizes, the most relevant perturbation which breaks the scale invariance is $\tilde{\lambda}:=|\lambda-\lambda_c|/\lambda_c$. This happens when $\ell^{-1}, T^{1/z}<\tilde{\lambda}^\nu$, where $\nu$ is the critical exponent of the correlation length, $\xi\sim\tilde{\lambda}^{-\nu}$. The scale invariance is then broken at the scale $\alpha\sim \xi\sim\tilde{\lambda}^{-\nu}$, and one finds,
\begin{align}\label{RCritical}
R\sim (\log{\tilde{\lambda}^{-\nu}})^{\tilde{\Delta}_R}\,\,\tilde{\lambda}^{\nu\Delta_R}.
\end{align}
As $\lambda$ gets closer to $\lambda_c$, and for sufficiently large system size, the next relevant perturbation becomes $\alpha\sim T^{-1/z}$. This happens at a cross-over regime when $\tilde{\lambda}^\nu\sim T^{1/z}>\ell^{-1}$, and the systems enters a thermally dominated regime, where
\begin{align}\label{RCriticalT}
R\sim (\log{\tilde{T}^{-1/z}})^{\tilde{\Delta}_R}\,\,T^{-\Delta_R/z}.
\end{align}
 
\subsection{High temperature limit}
In the high temperature regime, the quantum state converges to $\rho_T\propto\one$ and one expects the quantum properties of the state to be lost. We will show, indeed, that the parameter $R$ vanishes, and that this occurs with a universal scaling law of $R\sim 1/T$.
  
In the limit of very high temperature, i.e. when $\beta<<\beta_c\sim 1/\max_iE_i$, where $\max_iE_i$ is the largest energy eigenvalue, one can approximate $e^{-\beta E_i}\simeq 1 - \beta E_i$, thus finding from Eqs.~\eqref{JcP},\eqref{JqP} and~\eqref{UP} that
\begin{align}
J^c_{\mu \nu}&=\beta^2\Big(\sum_i p_i \partial_\mu E_i\partial_\nu E_i -\langle \partial_\mu H \rangle\langle \partial_\nu H \rangle\Big),\\
J^q_{\mu \nu}&\simeq\beta^2\sum_{ij}^{ E_i\neq E_j}(\partial_\mu H)_{ij}(\partial_\nu H)_{ji},\\
\mathcal{U}_{\mu \nu}&\simeq \frac{\beta^3}{4}\sum_{ij}^{ E_i\neq E_j} (E_i - E_j) (\partial_\mu H)_{ij}(\partial_\nu H)_{ji}.
\end{align}
Hence, 
\begin{align}\label{HighTR}
R:=||2i \mathcal{U}(J^c+J^q)^{-1}||_\infty \propto \beta,
\end{align}
which means that in the high temperature limit we expect the vanishing of $R$ with a universal $1/T$ rate.

\section{The quantum Ising chain in thermal equilibrium}
As an application of the above considerations, let's consider a paradigmatic model of spin-1/2 chains in thermal equilibrium, the one dimensional Ising model in transverse field. The model is defined by the Hamiltonian
\begin{align}
\HH(h)=-\frac{1}{2}\sum_{j=1}^M \left[\sx_j \sx_{j+1} + h \sz_j\right] .
\end{align}
We add an extra tuning parameter $\phi$, which corresponds to a rotation of all the spins
around the z-direction by $\phi/2$, obtaining
\begin{equation}
\label{HXYphi}
\HH(\phi,h) = g(\phi)\HH(h) g^\dag (\phi)\quad \text{ with }\quad g(\phi) =
 e^{\frac{i\phi}{2}\sum_i\sz_i}.
\end{equation}
The family of Hamiltonians parameterized by $\phi$ is clearly
isospectral, however the thermal state of the system does depend on $\phi$.
At $T=0$, the chain undergoes a quantum phase transition at $h=1$. For $h>1$, the system is in a disordered paramagnetic phase, with quasi-particle excitations given by spin-flips. For $h<1$, it is in an ordered phase, whose ground state is characterised by a long-range order $\lim_{r\to\infty}\left\langle \sx_0 \sx_r \right\rangle =(1-h^2)^{1/4}$.

The Hamiltonian  $\HH(\phi,h)$ can be reduced to an equivalent quasi-free fermionic model~\cite{Lieb1961}, under a Jordan-Wigner transformation,
$a_j=(\prod_{m<l}\sz_m)(\sx_j+i\sy_j)/2$ and a Fourier transform,
$d_k=\frac{1}{\sqrt{M}}\sum_{j} a_j e^{-i2 j k }$, with $k= \frac{n\pi}{M} $, $n\in \{-M/2,\dots M/2\}$, as
\begin{align}
\HH(\phi,h)&=\sum_k  \Psi_k H_k\Psi_k^\dag, \quad \textrm{ with }\quad H_k:=\frac{1}{2}\left( \begin{matrix}
\epsilon_k& i\Delta_k e^{i\phi}\\
-i\Delta_k e^{-i\phi}&-\epsilon_k
\end{matrix}\right)
\end{align}
where $\Psi_k:=(d_k,d_{-k}^\dag)^T$, $\Delta_k:= \sin k$, and $\epsilon_k:=(\cos{k}-h)$. A Bogoliubov transformation reduces the Hamiltonian to a diagonal form
\begin{align}
\HH= \sum_k \Lambda_k b_k^\dag b_k + const
\end{align}
where $b_k=d_k \cos\frac{\theta_k}{2}-id_{-k}^\dag
e^{i\phi}\sin\frac{\theta_k}{2}$, with $\theta_k:=\cos^{-1}(\epsilon_k/\Lambda_k)$, where
$\Lambda_k=\sqrt{\epsilon_k^2+\Delta_k^2}$.
The ground state of the model is the vacuum $b_k|Gs\rangle=0$ of the quasi-particle fermionic operators $b_k$. The thermal state can be expressed as
\begin{align}
\rho=\prod_k \frac{e^{-\beta \Lambda_k b_k^\dag b_k}}{Z_k}\qquad Z_k:=\Tr_k\,{e^{-\beta \Lambda_k b_k^\dag b_k}},
\end{align}
where $\Tr_k$ stands for the partial trace over the 2-dimensional Hilbert space of the Bogoliubov quasiparticle $b_k$. A thermal state of a quasi-free model is the prototypical example of Gaussian Fermionic state. By employing the formalism developed in~\cite{Carollo2018a} we can easily derive the SLD for the parameter estimation model $\{\lambda_\mu\}_{\mu=1}^3=\{\beta,h,\phi\}$
\begin{align}\label{SLD}
L_\mu&=\sum_k \Psi_k M^\mu_k\Psi_k^\dag - \frac{\eta_k^\mu}{2}\one, \quad \textrm{ with }\quad M_k^\mu:=\frac{1}{2}\vec{m}_k^\mu\cdot\vec{\sigma}
\end{align}
where
\begin{align}\label{SLDvec}
\left\{\begin{array}{l}
\vec{m}_k^\mu= -\frac{\partial \beta \Lambda_k}{\partial \lambda_mu} \hat{h}_k-\tanh{(\beta \Lambda_k)}\frac{\partial \hat{h}_k}{\partial \lambda_mu},\\
\eta_k^\mu=\frac{\partial \beta \Lambda_k}{\partial \lambda_mu} \tanh{(\beta \Lambda_k/2)},
\end{array}\right.\qquad
\hat{h}_k=\left(
\begin{array}{c}
\sin\theta_k \cos\phi\\
\sin\theta_k \sin\phi\\
\cos\theta_k
\end{array}
\right).
\end{align}
By plugging Eq.~\eqref{SLD} and Eq.~\eqref{SLDvec} into~\eqref{FI} and~\eqref{MUC},  one can then derive the Fisher information matrix $J$ and the MUC $\mathcal{U}$ as the sum of contributions of each quasi-momentum, i.e. $J=\sum_k J_k$ and $\mathcal{U}=\sum_k \mathcal{U}_k$
\begin{align}
J_k&=J_k^c+J_k^q\nonumber\\\label{JCMatrix}
J_k^c&=\frac{1}{4}(1-\tanh^2\beta\Lambda_k/2)\left( \begin{matrix}
\Lambda_k^2&-\beta\epsilon_k&0\\
-\beta\epsilon_k&\beta^2\epsilon_k^2/\Lambda_k^2&0\\
0&0&0
\end{matrix}\right)\\\label{JQMatrix}
J_k^q&=\frac{1}{2} \tanh(\beta\Lambda_k/2) \tanh(\beta\Lambda_k)\frac{\Delta_k^2}{\Lambda_k^3}\left( \begin{matrix}
0&0&0\\
0&1/\Lambda_k&0\\
0&0&\Lambda_k
\end{matrix}\right)\\\label{UMatrix}
\mathcal{U}_k&=\frac{1}{4}\tanh(\beta\Lambda_k/2) \tanh^2(\beta\Lambda_k)\frac{\Delta_k^2}{\Lambda_k^3}\left( \begin{matrix}
0&0&0\\
0&0&1\\
0&-1&0
\end{matrix}\right)
\end{align}
where $J^c=\sum_k J_k^c$ and $J^q=\sum_k J_k^q$ correspond to the "classical part" and "quantum part" of the QFIM, respectively.

It is convenient to deal with intensive quantities, by dividing the QFIM and the MUC by the size of the system, i.e. $J\to J/M$, $\mathcal{U}\to\mathcal{U}/M$. The above expressions may then be evaluated in the thermodynamic limit, where the summations are replaced by integrals, i.e. $J=\frac{1}{2\pi}\int_{-\pi}^\pi dk J_k $ and $\mathcal{U}=\frac{1}{2\pi}\int_{-\pi}^\pi dk \mathcal{U}_k$. The quantity $R$ can then be calculated from its definition in Eq.~\eqref{def:R}, by using the thermodynamic limit of the matrices $\mathcal{U}$ and $J$.
\begin{figure}[ht]
	\begin{subfigure}[b]{.5\textwidth}
		\centering
		\includegraphics[width=\linewidth]{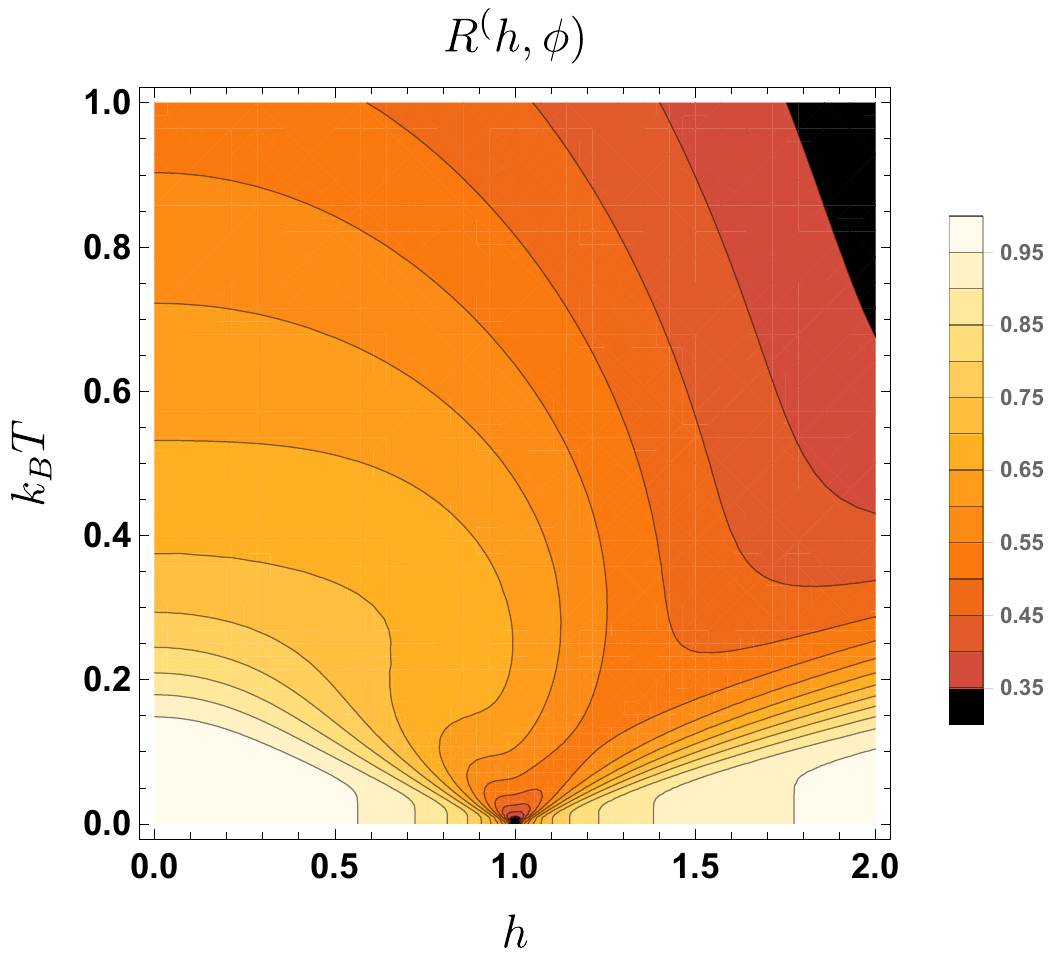}
		\label{ToverM}
	\end{subfigure}
	\hspace{.4cm}
	\begin{subfigure}[b]{.5\textwidth}
		\includegraphics[width=\linewidth]{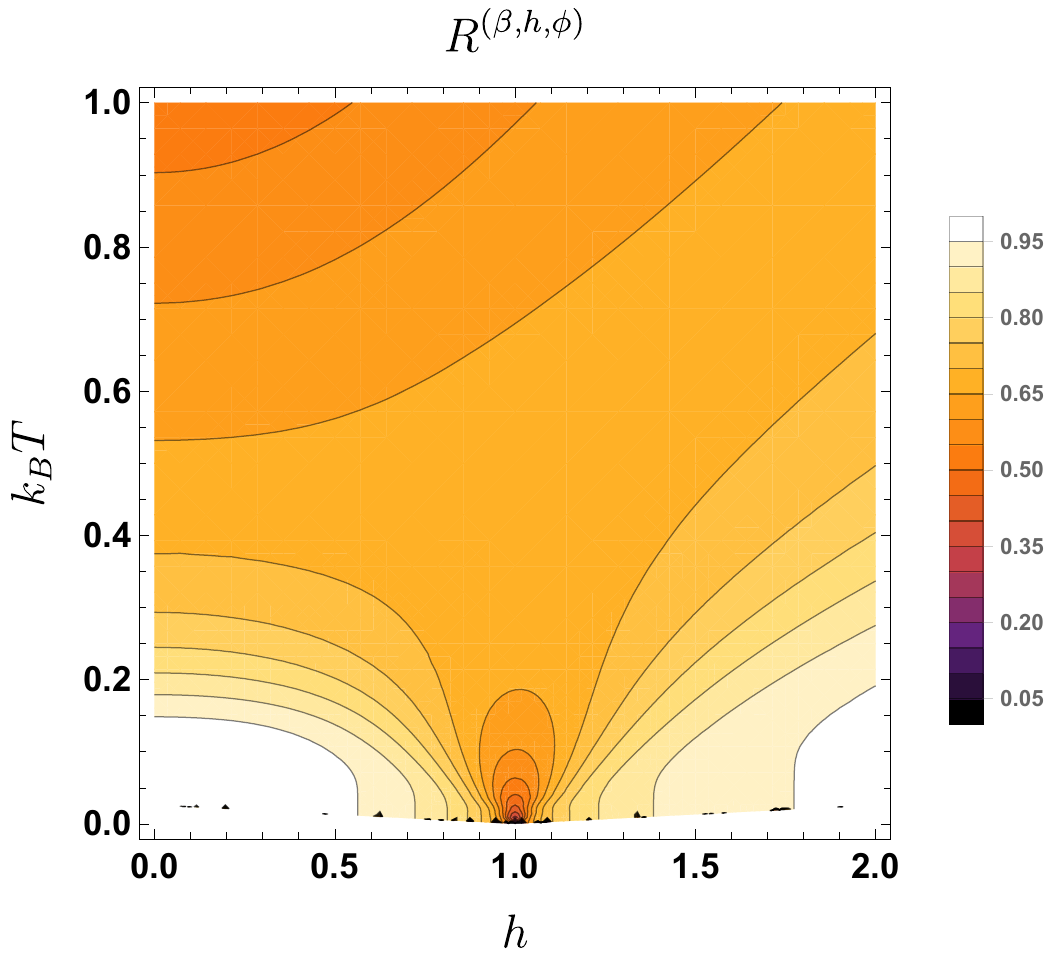}
		\label{DeltaoverT}
	\end{subfigure}
	\caption{Contour plot of $R:=||2i \mathcal{U}J^{-1}||_\infty$ in the plane $( h,T)$ of an Ising model in the thermodynamic limit, for two different parameter estimation schemes. On the left panel, $R^{( h,\phi)}$ for the two-parameter estimation $\mathcal{M}=( h,\phi)$. On the right panel, $R^{(\beta, h,\phi)}$ for the three-parameter estimation $\mathcal{M}=(\beta, h,\phi)$, where $\beta$ is the inverse temperature.}
	\label{Fig:R}
\end{figure}
In the left panel of Fig.~\ref{Fig:R}, we display the contour plot of $R^{( h,\phi)}$ as a function of $T$ and $ h$, in the case of the two-parameter estimation model, $\mathcal{M}=( h,\phi)$. Notice that the plot does not depend on $\phi$, since both the QFIM and MUC are independent of the specific value of $\phi$. This plot is obtained through the numerical integration of the Eqs.~\eqref{JCMatrix},~\eqref{JQMatrix} and ~\eqref{UMatrix}. Since the parameter-estimation model does not include $\beta$, the actual $J$ and $\mathcal{U}$ matrices considered in Fig.~\ref{Fig:R} are the $2\times 2$ matrices obtained from Eqs.~\eqref{JCMatrix},~\eqref{JQMatrix} and ~\eqref{UMatrix} by discarding the first rows and columns. An analogous result is displayed in the right panel of Fig.~\ref{Fig:R} for $R^{(\beta, h,\phi)}$, where the complete three-parameter estimation model, $\mathcal{M}=(\beta, h,\phi)$, has been considered. 

Both panels display the typical "V-shaped" phase diagram of quantum phase transitions, and one can recognise three distinctive regimes. The high temperature region, for $T\gg E_{max}/k_B$, where $E_{max}$ is the largest energy eigenvalue, the low temperature regime, for $T\lesssim\Delta/k_B$ and the quantum critical regime at $T\simeq 0$ and $ h\simeq h_c$. 
In the high temperature regime, for any fixed value of $h$, the parameter $R$ decreases asymptotically to zero, revealing that the quantum nature of the parameter estimation model is lost, due to thermal averaging. We will see in the next subsection that in this regime $R$ is expected to vanish with a universal $1/T$ rate.

In the low temperature regime $T\lesssim\Delta$, for values which are far from the critical value of the magnetic field, $h_c=1$, the parametric model shows its quantum incompatibility at its maximum. The phase diagram displays a plateau as function of temperature, and the quantumness of the system is nearly independent of $T$, showing a behaviour which is dictated by its zero temperature features.

Close to criticality, the system system displays a sharp minimum corresponding to $T=0$ and $ h=h_c=1$, where the value of $R$, in both parametric schemes drops abruptly to zero. This behaviour is due to the critical scaling of the Fisher Information which increases dramatically in the vicinity of a quantum criticality. The MUC diverges too in this regime, but with a slower rate compared to the QFIM. This causes the incompatibility condition to be relatively negligible in this regime and shows how the quantum multi-parameter scheme converges to a quasi-classical estimation problem. This implies that quantum nearly-critical systems, when used as probes in quantum estimation protocols, are quite beneficial not only for the dramatic enhancement of the sensitivity due to the divergent Fisher information, but also for the multi-parameter compatibility provided by the negligible value of $R$.

\subsection{Zero Temperature Limit and Quantum critical scaling}
\begin{figure}[ht]
	\begin{subfigure}[b]{.5\textwidth}
		\centering
		\includegraphics[width=\linewidth]{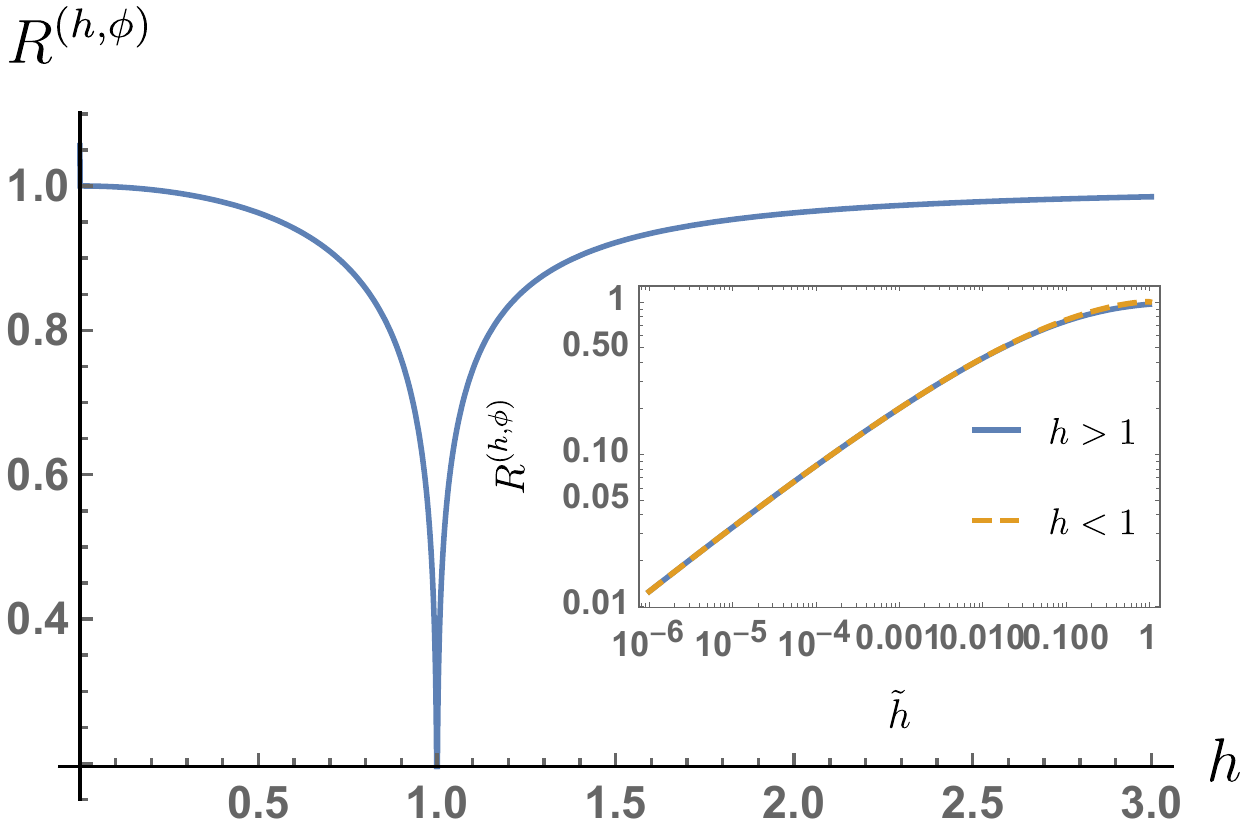}
		\caption{Zero temperature limit of $R$, as a function of the magnetic field $ h$. The parameter $R$ is clearly sensitive to the critical value of $ h=h_c=1$. The inset displays the critical dependence of $R$ on $\tilde{h}:=|h-h_c|/h_c$, in log-log scale.}
		\label{Roverl}
	\end{subfigure}
	\hspace{.4cm}
	\begin{subfigure}[b]{.5\textwidth}
		\includegraphics[width=\linewidth]{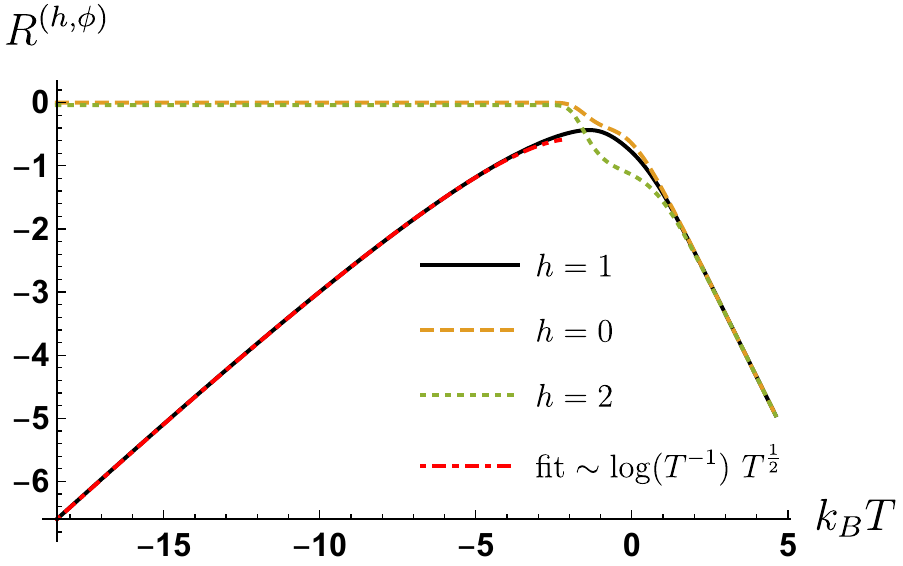}
		\caption{In log-log-scale, dependence of $R^{( h,\phi)}$ on temperature for three values of the magnetic field, $ h=0$ (anti-ferromagnetic phase), $ h=2$ (paramagnetic phase)  and $ h=1$ (critical value). The red dot-dashed line displays the scaling law $R=A\left(\log\left(\frac{8}{T} \right)-2\right)\,\, T^\frac{1}{2}$, with fitting parameter $A\simeq 0.74$.}
		\label{RoverT}
	\end{subfigure}
	\caption{}
\end{figure}

At zero temperature, and in the thermodynamic limit, one can calculate analytically the  quantities in Eqs.~\eqref{JCMatrix},~\eqref{JQMatrix} and~\eqref{UMatrix}, which read
\begin{align}
J_c&=0,\\
J^q&=\frac{f_q( h)}{4}\left( \begin{matrix}
0&0&0\\
0&\frac{1}{|1- h^2|}&0\\
0&0&1
\end{matrix}\right)\qquad\textrm{ with } \qquad f_q( h):=\left\{\begin{array}{ll}
1&| h|<1\\
\frac{1}{| h|^2}& | h|>1,
\end{array}\right.\\
\mathcal{U}_k&=\frac{g_q( h)}{4}\left( \begin{matrix}
0&0&0\\
0&0&1\\
0&-1&0
\end{matrix}\right)\qquad\textrm{ with } \qquad q_q( h):=\frac{(1+ h^2)K\left[\frac{4  h}{(1+ h)^2}\right] - (1+| h|)^2 E\left[\frac{4  h}{(1+ h)^2}\right]}
{\pi  h^2(1+| h|)},
\end{align}
where $K[x]$ and $E[x]$ are complete elliptic integrals of the first and second kind, respectively. This means that the parameter $R$ becomes
\begin{align}\label{RT0}
R(0)=2\sqrt{|1- h^2|}\frac{2g_q( h)}{f_q( h)}.
\end{align}
In Fig.~\ref{Roverl} we plot the zero temperature behaviour of $R$ as a function of the magnetic field $ h$. In the vicinity of the quantum phase transition $R$ displays a critical behaviour around $h_c=1$ which is compatible with the scaling law of Eq.~\eqref{RCritical}. Indeed, by expanding Eq.~\eqref{RT0} in series around $h_c$ one gets
\begin{align}\label{RT01}
R(0)=\frac{2\sqrt{2}}{\pi} \left(\log\left(\frac{8}{\tilde{ h}} \right)-2\right)\,\, \tilde{ h}^\frac{1}{2}+O\left(\tilde{ h}^\frac{3}{2}\right),
\end{align}
where $\tilde{ h}=| h-1|$. Hence, from the ansatz of Eq.~\eqref{RCritical} in subsection~\ref{sec:Critical}, we can infer that $\nu \Delta_{R}=1/2$. Since for the 1D quantum Ising model the correlation length critical exponent is $\nu=1$, we have
\begin{align}\label{RT02}
\Delta_R=-1/2\textrm{ and }\tilde{\Delta}_R=1.
\end{align}
According to the discussion of section~\ref{sec:Critical}, we expect a critical behaviour of $R$ also for $ h=h_c$ and $T>0$. In the critical region, as the temperature goes to zero, Eq.~\eqref{RCriticalT} predicts a scaling law with a multiplicative logarithmic correction, which for a dynamical critical exponent $z=1$ should be
\begin{align}\label{RCriticalT1}
R(T,\tilde{ h}=0)\sim  \log({T}^{-1})\,\,T^{\frac{1}{2}}.
\end{align}
In Fig.~\ref{RoverT} is displayed, in log-log scale, the dependence of $R^{( h,\phi)}$ on temperature for different values of $ h$. For non-critical values of $ h$ (orange-dashed and green-dotted lines) one recognises a flat behaviour corresponding to the low temperatures regimes, where the parameter $R$ reaches a plateau in which it saturates to its maximum value $R\sim1$. For the critical value of the magnetic field $\tilde{ h}=0$ (black solid line), one observes a scaling law which is compatible with Eq.~\eqref{RCriticalT1}. Finally, in the red dot-dashed line we display the fitting curve
\begin{align}\label{RTFit}
R(0)=A\left(\log\left(\frac{8}{T} \right)-2\right)\,\, T^\frac{1}{2},
\end{align} 
with a fitting parameter $A\simeq0.74$.
This confirms the universal behaviour of the parameter $R$ in the region close to criticality.

\subsection{High temperature limit}
In the high temperature limit, we can obtain the universal behaviour $1/T$ predicted by Eq.~\eqref{HighTR}. For high values of temperatures, i.e. for $\beta \max_i\Lambda_i<<1$, one can approximate $\tanh\beta\Lambda_k\simeq \beta\Lambda_k$. Thus one gets 
\begin{align}
J_k^c&\simeq\frac{1}{4}\left( \begin{matrix}
\Lambda_k^2&-\beta\epsilon_k&0\\
-\beta\epsilon_k&\beta^2\epsilon_k^2/\Lambda_k^2&0\\
0&0&0
\end{matrix}\right),\\
J_k^q&\simeq\frac{\beta^2}{4}\Delta_k^2\left( \begin{matrix}
0&0&0\\
0&1/\Lambda_k^2&0\\
0&0&1
\end{matrix}\right),\\
\mathcal{U}_k&\simeq\frac{\beta^3}{8}\Delta_k^2\left( \begin{matrix}
0&0&0\\
0&0&1\\
0&-1&0
\end{matrix}\right).
\end{align}
Notice that, considering the two parameter manifold $\mathcal{M}=( h,\phi)$, and restricting only to the quantum part of Quantum Fisher Information Matrix $J^q$, one gets
\begin{align}
R_q^{( h,\phi)}=||2i{(J^q)}^{-1} \mathcal{U}||_\infty= \sqrt{\frac{\det{2\mathcal{U}}}{\det{J^q}}}\simeq\beta\sqrt{\frac{\sum_k \Delta_k^2}{\sum_m \Delta_m^2/\Lambda_m^2 }}\xlongrightarrow{M\to\infty}\left\{\begin{array}{ll}
\beta&| h|>1\\
\beta| h|& | h|<1.
\end{array}\right.
\end{align}
Similarly, including the classical part of QFIM yields
\begin{align}\label{RhphiT}
R^{( h,\phi)}&=||2i{J}^{-1} \mathcal{U}||_\infty= \sqrt{\frac{\det{2\mathcal{U}}}{\det{(J^q+J^c)}}}\simeq \beta\sqrt{\frac{\sum_k\Delta_k^2}{M}}\xlongrightarrow{M\to\infty}\frac{\beta}{\sqrt{2}}. \end{align}
For the three parameter estimation model $\mathcal{M}=(\beta, h,\phi)$, we find the following value for the quantity $R$
\begin{align}
R^{(\beta, h,\phi)}=||2i{J^q}^{-1} \mathcal{U}||_\infty=\frac{2 |\mathcal{U}_{ h \phi}|}{\sqrt{ \left(1-\frac{J_{ h\beta}^2} {J_{\beta\beta} J_{ h h}}\right) J_{\phi \phi} J_{ h h} }}=\frac{R^{( h,\phi)}}{F^{(\beta, h)}},
\end{align}
where
\begin{align}
F^{(\beta, h)}:=&\sqrt{ \left(1-\frac{J_{ h\beta}^2} {J_{\beta\beta} J_{ h h}}\right)}\simeq\sqrt{ 1-\frac{(\sum_k \epsilon_k)^2}{\sum_m\Lambda_m^2}}\xlongrightarrow{M\to\infty}\frac{1}{\sqrt{1+ h^2}},
\end{align}
which leads to
\begin{align}\label{RhphibetaT}
R_k^{(\beta, h,\phi)}=\beta\sqrt{\frac{1+ h^2}{2}}.
\end{align}
Eqs.~\eqref{RhphiT} and~\eqref{RhphibetaT} confirm the universal behaviour of $R\propto1/T$, in the high temperature limit, predicted by Eq.~\eqref{HighTR}. In Fig.~\ref{RoverT} we show the dependence of $R^{( h,\phi)}$ on temperature, for different values of the magnetic field $ h$. Note that in the high temperature regime, independently of $ h$, the plot displays the predicted $1/T$ universal scaling law of $R^{( h,\phi)}$.
\section{Conclusion and outlook}
We have introduced a novel approach to quantitatively assess the ``quantumness'' of a parameter estimation model. To this end, we resorted to the idea of incompatibility 
of parameters in a quantum estimation model. The crucial concept is the mean Uhlmann curvature $\mathcal{U}$, a quantity which plays a pivotal role in the description 
of the geometric and topological features of the space of the density matrices. In multi-parameter quantum metrology, the MUC accounts for the compatibility condition, 
i.e. a prescription which guarantees whether or not the Cram\'er-Rao bound can be saturated. In this work we defined the ratio $R$, which is the largest eigenvalue of the 
matrix $2iJ^{-1}\mathcal{U}$, that, roughly speaking, quantifies the relative size between MUC and Quantum Fisher information. We argued that $R$ provides a good 
quantitative measure of the discrepancy between an inherently quantum and a quasi-classical multi-parameter estimation problem. We demonstrated that this quantity is 
a real number lying in the interval $\{0,1\}$ and that the two limiting values correspond to the two extreme cases: $R=0$ signals the quasi-classical case, where independent 
parameters can be simultaneously estimated, reaching the same accuracy as the individual estimation strategy; $R=1$ flags the fully quantum case, where the indeterminacy 
arising from the uncertainty principle hinders the accuracy of the parameter estimation, in a way which cannot be neglected, not even in the limit of infinite copies.\\
\indent We envision the possibility of using this tool to investigate the metrological features of  many-body quantum systems. Indeed, it is known that in quantum parameter estimation 
theory peculiar quantum many-body states may be exploited as a probe to enhance the accuracy of estimation protocols~\cite{Zanardi2008,Braun2018}. On the other hand, 
one may think of using methods developed in quantum metrology to investigate and characterise many-body systems. 
We have applied the above idea to a paradigmatic example of a quantum many-body system and have investigated in depth different regions of the phase diagram from the perspective of the parameter $R$. By exployiting this paradigmatic model, one can draw general conclusions of the generic behaviour of quantum many body systems.
We have also performed a scaling analysis on $R$ as a function of both temperature and magnetic field.

Quantum critical systems are ideal candidates to use 
these tools, since quantum parameter estimation provides a novel operational approach to investigate
equilibrium~\cite{Carollo2005,Zhu2006,Hamma2006,Zanardi2006,CamposVenuti2007,CamposVenuti2008,Zanardi2007,Zanardi2007a,Garnerone2009a,Rezakhani2010}
and out-of-equilibrium~\cite{Magazzu2015,Magazzu2016,Guarcello2015,Spagnolo2015,Spagnolo2017,Spagnolo2018,Valenti2018,Spagnolo2018a}
quantum critical phenomena~\cite{Banchi2014,Marzolino2014,Kolodrubetz2013,Carollo2018,Carollo2018a,Marzolino2017}. The concept developed in this work indeed may shed new 
light on the nature of correlations and the interplay between competing orders both in equilibrium and non-equilibrium quantum critical phenomena.

\section*{Acknowledgments}
We would like to thank Francesco Albarelli and Animesh Datta  to spot an error in Eq.~\eqref{CH} and point towards a workaround~\cite{Albarelli2019}.
This work was supported by the Grant of the Government of the Russian Federation, contract No. 074-02-2018-330 (2). 
We acknowledge also partial support by Ministry of Education, University and Research of the Italian Government.

\vspace{180pt}

\providecommand{\newblock}{}
\bibliographystyle{iopart-num}
\bibliography{ref}
\end{document}